\documentclass[10pt,twocolumn,a4paper,conference]{IEEEtran}
\usepackage{flushend} 
\usepackage{graphicx}
\usepackage{epsfig}
\usepackage{latexsym}
\usepackage{amsfonts}
\usepackage{here}
\usepackage{rawfonts}
\usepackage[latin1]{inputenc}
\usepackage[T1]{fontenc}
\usepackage{calc}
\usepackage{url}
\usepackage{enumerate}
\usepackage{color}
\usepackage{amssymb}
\usepackage{bm}
\usepackage{upref}
\usepackage{times}
\usepackage{amsmath}
\usepackage{cite}

\newcommand{\ms}{\;\;}

\newcommand{\mycaption}[1]{\vspace*{-1.6em}\caption{#1}\vspace*{-1.0em}}

\newcommand{\qed}{\nobreak \ifvmode \relax \else
  \ifdim\lastskip<1.5em \hskip-\lastskip
  \hskip1.5em plus0em minus0.5em \fi \nobreak
  \vrule height0.75em width0.5em depth0.25em\fi}

\newlength{\aligntop}
\newlength{\alignbot}
\makeatletter
\renewenvironment{align}{%
  \vspace{\aligntop}
  \start@align\@ne\st@rredfalse\m@ne
}{%
  \math@cr \black@\totwidth@
  \egroup
  \ifingather@
    \restorealignstate@
    \egroup
    \nonumber
    \ifnum0=`{\fi\iffalse}\fi
  \else
    $$%
  \fi
  \ignorespacesafterend%
  \vspace{\alignbot}\par\noindent
}
\makeatother

\IEEEoverridecommandlockouts

\author{Manav R. Bhatnagar\\ 
 Department of Electrical Engineering\\ Indian Institute of Technology Delhi\\ Hauz Khas, New Delhi, India, IN-110016\\
    Email: \protect\url{manav@ee.iitd.ac.in} 
\vspace*{-1.5em}%
}\date{}

\title{On the Capacity of CSI Based Transmission Link Selection in Decode-and-Forward Cooperative System \vspace*{-0.28em}}

\begin{document}
\IEEEoverridecommandlockouts
\renewcommand{\baselinestretch}{1.0}
\maketitle
\vspace*{-1.0em}
\begin{abstract}
In this paper, we study the problem of best transmission link selection in a decode-and-forward (DF) cooperative system from capacity point of view. The transmission link can be a cooperative (via a relay) or direct link between the source and destination nodes.
In a two-hop DF system with multiple relays and a direct link in between the source and destination, the transmission link selection can be performed based 
on full or partial channel state information (CSI) of all links involved in cooperation. We derive analytical ergodic capacity of full and partial CSI
based path selection schemes in the DF cooperative system. Further, the full and partial CSI based link selection schemes are compared with help of these expressions. 
\end{abstract}\vspace*{-0.5em}
\vspace*{0.0em}  
\section{Introduction}\vspace*{-0em}
Virtual antenna system can be created by exploiting cooperative diversity among different cooperating nodes~\cite{lanem03}. The cooperation may be regenerative (DF)~\cite{bhatn11,bhatn12a,bansa13,bansa13a} or non regenerative (amplify-and-forward (AF))~\cite{maham09,maham09a,maham12} relaying technique. The AF protocol can use very simple relay processing of merely scaling the received signal at the relay with a fixed gain~\cite{mk12,mk13b,mk13c} but it has a problem of noise amplification~\cite{bhatn11,bhatn12a,bansa13,bansa13a}. The DF relaying is more practical than AF relaying because of its digital processing nature~\cite{bhatn11,bhatn12a,bansa13,bansa13a}. Depending upon the source-relay channel, the relay can demodulate the data correct or wrong; 
the erroneous relaying of the data causes significant penalty in term of average error performance of the destination receiver~\cite{lanem03,bhatn11,bhatn12a,bansa13,bansa13a}. In~\cite{bhatn11,bhatn12a,bansa13,bansa13a}, maximum likelihood decoders have been discussed for avoiding the error propagation in DF networks due to the erroneous relaying of a practical relay node. 
Relay selection is another useful method for improving the quality of relaying and data-rate of the cooperative system~\cite{blets06}, where a best relay is selected for cooperation based on full~\cite{goert09} or partial~\cite{bhatn13d,beres08} CSI of the links involved in the cooperation.  

Finding the ergodic capacity of the DF networks is an important problem~\cite{bhatn13c}. In~\cite{lee09}, the ergodic capacity of different opportunistic DF relaying
schemes over dual-hop Rayleigh fading channels is derived without considering a direct link between the source and the destination. The ergodic capacity of relay selection in \emph{repetition} based two-hop DF cooperative system with a direct link in symmetric source-to-relay and symmetric relay-to-destination links is derived in~\cite{nikja09}. The ergodic capacity of a partial CSI based relay selection scheme in DF cooperative system without a direct link is derived in~\cite{rui10}. In~\cite{worad08}, the outage capacity was derived for several dual-hop DF schemes, including opportunistic relaying in DF system with a direct link, a smart selective DF scheme, and a hybrid system. It is argued in~\cite{worad08} that the outage capacity can be improved by using a smart selection scheme that can revert to non-cooperation mode if the direct link is better than the cooperative links. 
The average error performance of the full CSI based selection combining in two-hop DF cooperative system with multiple relays and a direct link is performed in~\cite{selva12}, where the best link (cooperative or direct) between the source and destination is chosen on the basis of the instantaneous signal-to-noise ratios (SNRs) of all links. A general analytical framework for link selection in beamforming and combining based DF networks where source, a single relay, and destination are equipped with multiple antennas, is presented in~\cite{bhatn13b}.

In this paper, we derive the ergodic capacity of full and partial CSI based link selection scheme in a dual-hop DF cooperative system with multiple relays and a direct link between the source and destination node. Both schemes are compared on the basis of the derived capacity expressions. 
\begin{figure}[t!]\vspace*{-0.0em}
  \begin{center}\hspace*{0em}
    \psfig{figure=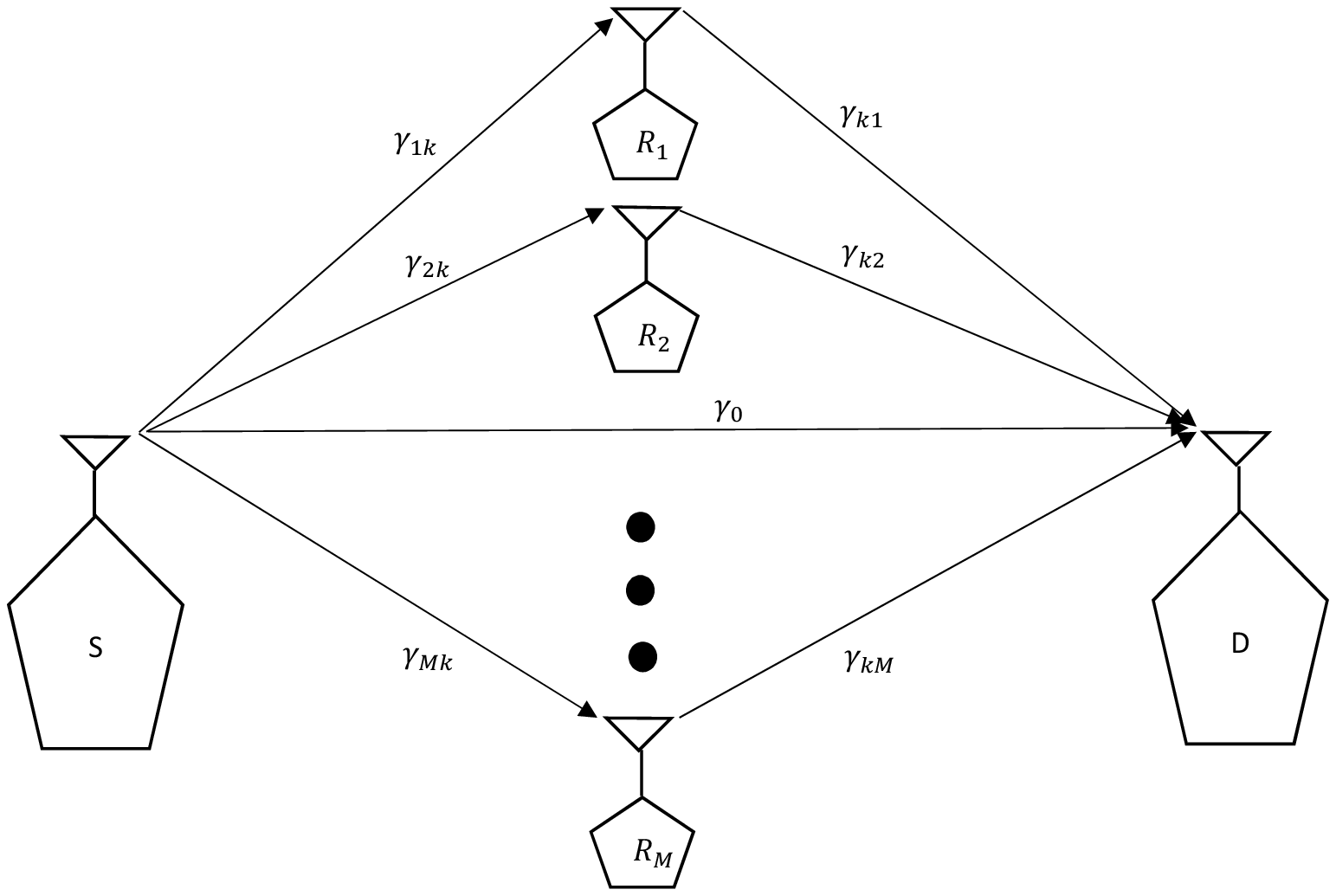,height=2.0in,width=3.25in}
    \vspace*{2em}
    \mycaption{Two-hop cooperative communication system with $M$ relays and a direct link between S and D; $\gamma_{ik}$ denotes the instantaneous SNR of the S-R$_i$ link; $\gamma_{ki}$ denotes the instantaneous SNR of the R$_i$-D link; $\gamma_{0}$ denotes the instantaneous SNR of the S-D link.}
    \label{fig:cpbd}
    \vspace*{-0.0em}
  \end{center}
\end{figure}  
\section{System Model}
Let us consider a two-hop cooperative system with a source (S), one destination (D), and $M$ regenerative relays (R$_1$,R$_2$,..,R$_M$); each node is equipped with a single antenna, as shown in Fig.~\ref{fig:cpbd}. It is assumed that a node can either transmit or receive the data at a time. There exists a direct link along with $M$ two-hop cooperative links between S and D. All links are assumed to be independent but non-identically distributed (i.n.i.d.) Rayleigh fading with their instantaneous SNRs complying to the Exponential distribution as follows: $\gamma_{ik}\sim \mathcal{E}\left(\bar{\gamma}_{ik}\right), \gamma_{ki}\sim \mathcal{E}\left(\bar{\gamma}_{ki}\right)$, and $\gamma_{0}\sim \mathcal{E}\left(\bar{\gamma}_{0}\right)$, 
where $\mathcal{E}\left(m\right)$ denotes the Exponential distribution with mean $m$; $\gamma_{ik}$, $\gamma_{ki}$, and $\gamma_{0}$ denote the instantaneous SNRs and $\bar{\gamma}_{ik}$, $\bar{\gamma}_{ki}$, and $\bar{\gamma}_{0}$ represent the average SNRs of the S-R$_{i}$, R$_{i}$-D, and S-D links, respectively, and $i=1,2,..,M$. Let us describe the considered link selection schemes in the following subsections. 
\subsection{Full CSI Based Link Selection}
In the full CSI based transmission link selection scheme, 
the best diversity branch for transmission of the data of the source is selected as follows~\cite{selva12}:
\begin{align}
\label{eq:condfullcsi}
\begin{array}{ccc}
&\text{if}\:\gamma_j>\gamma_i\: \text{and} \:\gamma_j>\gamma_0, &\!\!\!\! \text{then select}\:S-R_j-D\: \text{link}.\\
&\hspace*{-4.2em}\text{if}\:\gamma_0>=\gamma_i,&\hspace*{-3em} \text{then select}\: S-D\: \text{link},
\end{array}
\end{align}
where $j\neq i$, $j=1,2,...,M$, and $\gamma_i=\min\left\{\gamma_{ik},\gamma_{ki}\right\}$ whose cumulative distribution function (CDF) 
can be obtained as follows:
\begin{align}
\label{eq:CDF_max}
F_{\gamma_i}(\gamma)&=1-(1-F_{\gamma_{ik}}(\gamma))(1-F_{\gamma_{ki}}(\gamma))\nonumber\\
&=1-e^{-\gamma/\bar{\gamma}_i},
\end{align}
where $F_X(x)$ denotes the CDF of $X$ and $1/\bar{\gamma}_i=1/\bar{\gamma}_{ik}+1/\bar{\gamma}_{ik}$.
\subsection{Partial CSI Based Link Selection}
With the knowledge of the channels of R$_i$-D and S-D links, the best transmission link is selected as follows:
\begin{align}
\label{eq:condpartcsi}
\begin{array}{ccc}
&\hspace*{-1em}\text{if}\:\gamma_{kj}>\gamma_{ki}\: \text{and} \:\gamma_{kj}>\gamma_0, & \!\!\!\!\!\text{then select}\:S-R_j-D\: \text{link},\\
&\hspace*{-6.3em}\text{if}\:\gamma_0>=\gamma_{ki},& \hspace*{-3.2em}\text{then select}\: S-D\: \text{link},
\end{array}
\end{align}
where $j\neq i$. 
The partial CSI based link selection scheme in \eqref{eq:condpartcsi} is influenced by the relay selection scheme given in~\cite{beres08};
it is easier to implement this scheme in practice.  
\vspace*{0em}
\section{Capacity of Full CSI Based Transmission Link Selection Scheme}
If a cooperative link is selected, 
then the communication from S to D
takes place in two orthogonal channels via a DF relay (R$_i$).
Therefore, both hops in a cooperative link act independently in the sense
that decoding/encoding is performed at the intermediate relay (R$_i$).
Thus, a DF cooperative link is equivalent
to a series network, which means that the capacity of the
link is dominated by the worst hop. 
Since the capacity is a
monotonous function of SNR, the minimum of the capacities
of S-R$_i$ and R$_i$-D hops equals 
the capacity of the weakest of the two hops. 
Therefore, the equivalent SNR of the cooperative link, from a capacity
point of view, is given by $\gamma_i$~\cite{bhatn13c}. Hence, 
the capacity of the $i$-th cooperative link in the considered DF system is given as
\begin{align}
\label{eq:cap_coop}
C_{\text{COP}}\left(\gamma_i\right)=\frac{1}{2}\text{log}_2\left(1+\gamma_i\right),
\end{align}
where the factor of $1/2$ is included because of repeated transmission of the data over cooperative link via the relay. Further, the capacity of the direct link between the source and destination is given as
\begin{align}
\label{eq:cap_dir}
C_{\text{D}}\left(\gamma_0\right)=\text{log}_2\left(1+\gamma_0\right).
\end{align}
It can be noticed that in the full CSI based scheme, a transmission link between the source and destination that contains best instantaneous SNR is selected for transmission of the source's data to the destination. Let us denote the best instantaneous SNR by $\gamma_b$, then the instantaneous capacity of the full CSI combining based DF system can be written as
\begin{align}
\label{eq:cap_coop_tot}
&C_F\left(\gamma_i,\gamma_0\right)\!=\!\!\sum^M_{i=1}\mbox{Pr}\left({\gamma_b=\gamma_i}\right)C_{\text{COP}}\left(\gamma_i|\gamma_b=\gamma_i\right)\nonumber\\
&\hspace*{4em}+\mbox{Pr}\left({\gamma_b=\gamma_0}\right)C_{\text{D}}\left(\gamma_0|\gamma_b=\gamma_0\right),
\end{align}
where  $\mbox{Pr}\left({\cdot}\right)$ denotes the probability. It can be seen from (\ref{eq:cap_coop_tot}) that the averge conditional capacity depends upon the capacity of all links. 

It can be deduced from \eqref{eq:condfullcsi} that a cooperative link (S-R$_i$-D link) is chosen if its minimum instantaneous SNR is greater than the minimum instantaneous SNRs of all other cooperative links and instantaneous SNR of the direct link (S-D link). We can alternately say that the $i-$th cooperative link will be selected for transmission if its minimum instantaneous SNR, i.e., $\gamma_i$, is greater than the maximum of the minimum instantaneous SNRs of the remaining $M-1$ cooperative links and than the instantaneous SNR of the direct link. Moreover, the direct link will be selected if its instantaneous SNR is greater than the maximum of the minimum instantaneous SNRs of all cooperative links. Let $\gamma_b$ denotes the best instantaneous SNR obtained from \eqref{eq:condfullcsi}, we can express the aforementioned conditions mathematically as
\begin{align}
\label{eq:cond}
\begin{array}{ccc}
\text{if}\:\gamma_b=\gamma_i,& \text{then}& \gamma_i>\gamma^{(i-)}_M \:\text{and} \ms \gamma_i>\gamma_0,\\
\text{if}\:\gamma_b=\gamma_0,& \text{then}& \hspace*{-5.87em}\gamma_0>\gamma_M,
\end{array}
\end{align}
where $
\gamma_M=\max\left\{\gamma_1,\gamma_2,...,\gamma_M\right\}$ 
and $\gamma^{(i-)}_M=\max\left\{\gamma_1,\gamma_2,...,\gamma_{i-1},\gamma_{i+1},...,\gamma_M\right\}$.


By using~\eqref{eq:CDF_max} and~\cite[Eq.~(6.78)]{papou02}, it can be shown that the CDF of $\gamma_M$ will be
\begin{align}
\label{eq:CDFmax1}
&F_{\gamma_M}(\gamma)=\Pi^M_{i=1}F_{\gamma_i}(\gamma)\nonumber\\
&=1-\sum^M_{j_1=1}e^{-\gamma/\bar{\gamma}_{j_1}}+\underset{j_1>j_2}{\sum^M_{j_1=1}\sum^M_{j_2=1}}e^{-\gamma\left(1/\bar{\gamma}_{j_1}+1/\bar{\gamma}_{j_2}\right)}-...\nonumber\\
&+(-1)^{M}\underset{j_1>j_2>..>j_M}{\sum^M_{j_1=1}\sum^M_{j_2=1}\cdots\sum^M_{j_M=1}}e^{-\gamma\left(1/\bar{\gamma}_{j_1}+1/\bar{\gamma}_{j_2}+...+1/\bar{\gamma}_{j_M}\right)}\nonumber\\
&=1+\sum^M_{m=1}(-1)^m\mathbb{S}_me^{-\gamma/\bar{\gamma}'_m},
\end{align}
where
$\mathbb{S}_m=\underset{j_1>j_2>..>j_m}{\sum^M_{j_1=1}\sum^M_{j_2=1}\cdots\sum^M_{j_m=1}}$
and $1/\bar{\gamma}'_m=\sum^m_{l=1}(1/\bar{\gamma}_{j_l})$. Similarly, 
it can be shown that
\begin{align}
\label{eq:CDFmax2}
&F_{\gamma^{(i-)}_M}(\gamma)
=1+\sum^{M-1}_{m=1}(-1)^m\mathbb{S}^{(i-)}_me^{-\gamma/\bar{\gamma}'_m},
\end{align}
where
$\mathbb{S}^{(i-)}_m=\underset{j_1>j_2>..>j_m}{\sum^M_{j_1=1\atop j_1\neq i}\sum^M_{j_2=1\atop j_2\neq i}\cdots\sum^M_{j_m=1\atop j_m\neq i}}$. 

By using \eqref{eq:cond}, we can marginalize \eqref{eq:cap_coop_tot} over different channels as follows:
\begin{align}
\label{eq:erg_cap}
C_F&=\frac{1}{2}\sum^M_{i=1}\Big(\int^\infty_{0}\text{log}_2\left(1+x\right)f_{\gamma_i}(x)\int^x_0f_{\gamma^{(i-)}_M}(y)\nonumber\\
&\times \int^x_0f_{\gamma_0}(z)dz dy dx\Big)+\int^\infty_{0}\text{log}_2\left(1+x\right)f_{\gamma_0}(x)\nonumber\\
&\times\int^x_0f_{\gamma_M}(y)dy dx.
\end{align}
It is shown in Appendix~\ref{app:1} that the ergodic capacity of the full CSI based best link selection scheme will be
\begin{align}
\label{eq:CFCSI}
C_F&= \frac{1}{2}\sum^M_{i=1}\Bigg(e^{1/\bar{\gamma}_i}\text{log}_2e E_1\left(\frac{1}{\bar{\gamma}_i}\right)-\frac{\text{log}_2e}{\bar{\gamma}_i}\sum^{M-1}_{m=0}(-1)^m\nonumber\\
&\times\mathbb{S}^{(i-)}_{0,m}\left(\frac{1}{\bar{\gamma}_i}+\frac{1}{\bar{\gamma}''_m}\right)^{-1} e^{1/\bar{\gamma}_i+1/\bar{\gamma}''_m}E_1\left(\frac{1}{\bar{\gamma}_i}+\frac{1}{\bar{\gamma}''_m}\right)\Bigg)\nonumber\\
&+e^{1/\bar{\gamma}_0}\text{log}_2e E_1\left(\frac{1}{\bar{\gamma}_0}\right)+\frac{\text{log}_2e}{\bar{\gamma}_0}\sum^{M}_{m=1}(-1)^m\mathbb{S}_{m}\nonumber\\
&\times\left(\frac{1}{\bar{\gamma}_0}+\frac{1}{\bar{\gamma}'_m}\right)^{-1} e^{1/\bar{\gamma}_0+1/\bar{\gamma}'_m}E_1\left(\frac{1}{\bar{\gamma}_0}+\frac{1}{\bar{\gamma}'_m}\right), 
\end{align}
where $\mathbb{S}^{(i-)}_{0,m}=\underset{j_0>j_1>..>j_m}{\sum^M_{j_0=0\atop j_0\neq i}\sum^M_{j_1=0\atop j_1\neq i}\cdots\sum^M_{j_m=0\atop j_m\neq i}}$, 
$1/\bar{\gamma}''_m=\sum^m_{l=0}(1/\bar{\gamma}_{j_l})$, and $E_1\left(\mu\right)=\int^\infty_1t^{-1}e^{-t\mu} dt$ is the Exponential-integral function with an alternative form given in~\cite[Eq. (8.211.1)]{grand07}.\vspace*{0em}
\section{Capacity of Partial CSI Based Transmission Link Section Scheme}
For simplicity, let us define new random variables (RVs) as $\Omega_0\triangleq\gamma_0, \Omega_i\triangleq\gamma_{ki}$, $\bar{\Omega}_0\triangleq\bar{\gamma}_0$, and $\bar{\Omega}_i\triangleq\bar{\gamma}_{ki}, i=1,2,...,M$. In the partial CSI based selection scheme, if the $i$-th cooperative link is selected for transmission of the data from S to D, then $\Omega_i$, which is chosen as the best instantaneous SNR according to \eqref{eq:condpartcsi}, can be less or greater than $\gamma_{ik}$. Therefore, it can be deduced from \eqref{eq:condpartcsi} that 
\begin{align}
\label{eq:condpartsel}
\begin{array}{ccc}
\text{if}\:\gamma_b=\Omega_i,& \text{then}& \Omega_{i}>\Omega^{(i-)}_M \:\text{and}\:\: \Omega_i<\gamma_{ik}\\
& \text{or} &\Omega_{i}>\Omega^{(i-)}_M\: \text{and}\:\: \Omega_i>\gamma_{ik}, \\
\text{if}\:\gamma_b=\Omega_0,& \text{then}&\hspace*{-5.45em} \Omega_0>\Omega^{(0-)}_M,
\end{array}
\end{align}
where $\Omega^{(l-)}_M, l=0,1,...,M$ is a RV given as
\begin{align}
\label{eq:max2}
\Omega^{(l-)}_M\triangleq\max\left\{\Omega_0,\Omega_1,...,\Omega_{l-1},\Omega_{l+1},...,\Omega_M\right\}.
\end{align}
By following the similar procedure given in \eqref{eq:CDFmax1}, it can be shown that
\begin{align}
\label{eq:cdf_omegaM}
&F_{\Omega^{(l-)}_M}(\gamma)=1-\sum^{M-1}_{m=0}(-1)^m\mathbb{S}^{(l-)}_{0,m}e^{-\gamma/\bar{\Omega}'_m},
\end{align}
where $1/\bar{\Omega}'_m=\sum^m_{t=0}(1/\bar{\Omega}_{j_t})$. The instantaneous capacity of partial CSI based selection scheme can be written as
\begin{align}
\label{eq:cap_coop_tot_partial_rel_sel}
C_P\left(\gamma_i,\gamma_0\right)=&\sum^M_{i=1}C_{\text{COP}}\left(\gamma_i|\gamma_b=\Omega_i\right)
+C_{\text{D}}\left(\Omega_0|\gamma_b=\Omega_0\right).
\end{align}
From \eqref{eq:condpartsel} and \eqref{eq:cap_coop_tot_partial_rel_sel}, the ergodic capacity of partial CSI based selection scheme can be expressed in the integral form as
\begin{align}
\label{eq:erg_cap_part_rel_sel}
&C_P=\frac{1}{2}\sum^M_{i=1}\Bigg(\int^\infty_{0}\text{log}_2\left(1+x\right)f_{\gamma_{ik}}(x)\int^\infty_x f_{\Omega_i}(y)\nonumber\\
&\times\int^y_0f_{\Omega^{(i-)}_M}(z)dz dy dx+\int^\infty_{0}\text{log}_2\left(1+x\right)f_{\Omega_{i}}(x)\nonumber\\
&\times \int^x_0 f_{\Omega^{(i-)}_M}(y)\int^\infty_xf_{\gamma_{ik}}(z)dz dy dx\Bigg)\nonumber\\
&+\int^\infty_{0}\text{log}_2\left(1+x\right)f_{\Omega_{0}}(x)\int^x_0 f_{\Omega^{(0-)}_M}(y)dy dx.
\end{align}
It is shown in Appendix~\ref{app:2} that the ergodic capacity of the partial CSI based path selection scheme in the DF relaying will be
\begin{align}
\label{eq:erg_cap_part_rel_sel1}
&C_P=\frac{1}{2}\sum^M_{i=1}\Bigg[e^{1/\bar{\gamma}_i}\text{log}_2e E_1\left(\frac{1}{\bar{\gamma}_i}\right)-\frac{\text{log}_2e}{\bar{\gamma}_{ki}}\sum^{M-1}_{m=0}(-1)^m\nonumber\\
&\times\mathbb{S}^{(i-)}_{0,m}\left(\frac{1}{\bar{\gamma}_{ik}}\left(\frac{1}{\bar{\gamma}_i}+\frac{1}{\bar{\Omega}'_m}\right)^{-1}+1\right)\left(\frac{1}{\bar{\gamma}_{ki}}+\frac{1}{\bar{\Omega}'_m}\right)^{-1} \nonumber\\
&\times e^{1/\bar{\gamma}_i+1/\bar{\Omega}'_m} E_1\left(\frac{1}{\bar{\gamma}_i}+\frac{1}{\bar{\Omega}'_m}\right)\Bigg]+e^{1/\bar{\gamma}_0}\text{log}_2e E_1\left(\frac{1}{\bar{\gamma}_0}\right)\nonumber\\
&-\frac{\text{log}_2e}{\bar{\gamma}_0}\sum^{M-1}_{m=0}(-1)^m\mathbb{S}^{(0-)}_{0,m}\left(\frac{1}{\bar{\gamma}_0}+\frac{1}{\bar{\Omega}'_m}\right)^{-1} e^{1/\bar{\gamma}_0+1/\bar{\Omega}'_m}\nonumber\\
&\times E_1\left(\frac{1}{\bar{\gamma}_0}+\frac{1}{\bar{\Omega}'_m}\right).
\end{align}
\section{Comparisons and Numerical Results}
\subsection{Comparison of Partial and Full CSI Based Link Selection Schemes in I.I.D. Links}
\begin{figure}[t!]\vspace*{-1.0em}
  \begin{center}\hspace*{-1.8em}
    \psfig{figure=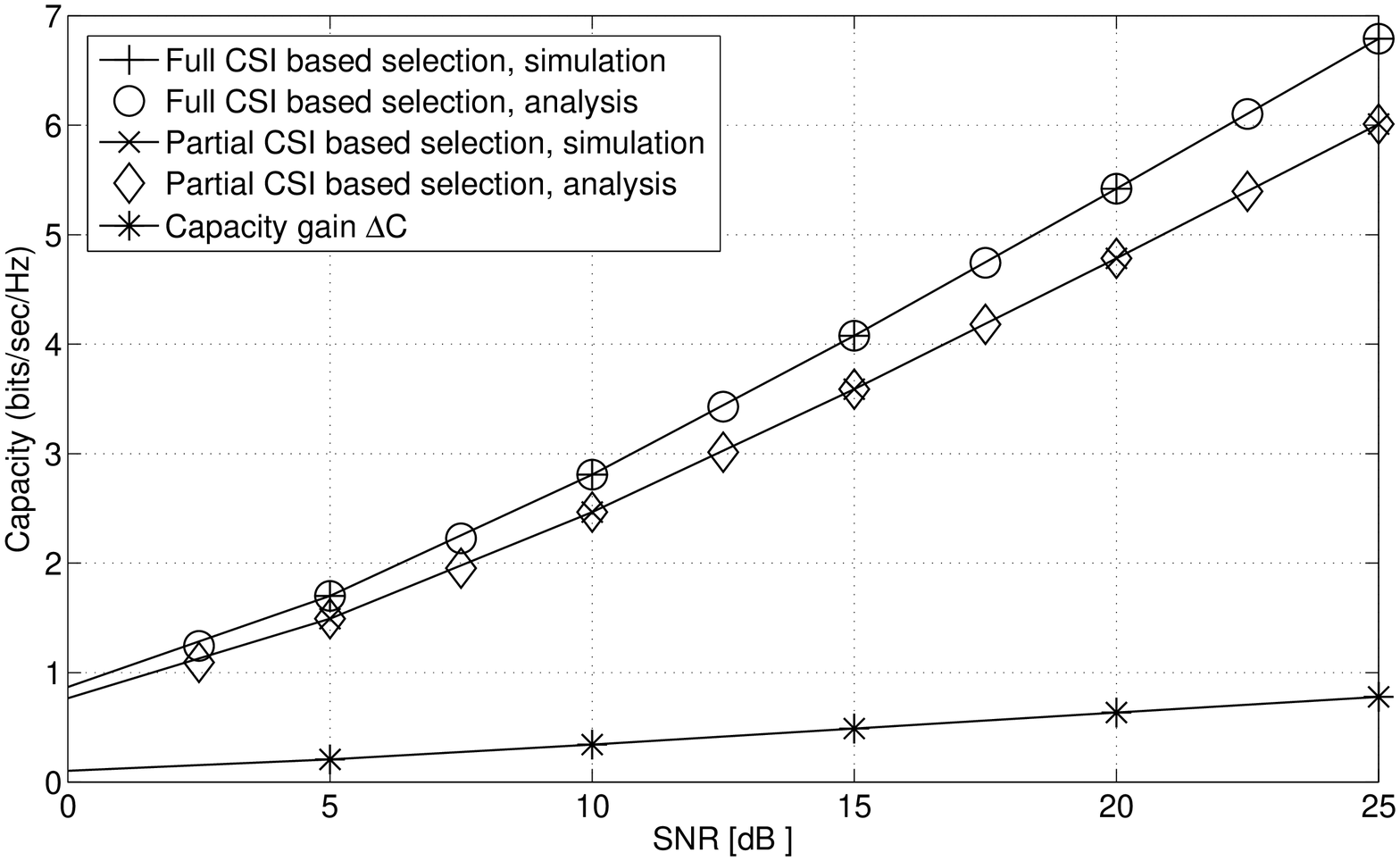,height=3.2in,width=4.0in}
    \vspace*{0em}
    \mycaption{Analytical and simulated ergodic capacity of full and partial CSI based link selection schemes in the two-hop DF system with a single relay and i.i.d. channels.}
    \label{fig:ray2}
    \vspace*{-0.0em}
  \end{center}
\end{figure}

For independent and identically distributed (i.i.d.) links and $M=1$, i.e., a DF cooperative system with a single relay and $\bar{\gamma}_0=\bar{\gamma}_{1k}=\bar{\gamma}_{k1}=\bar{\gamma}$,
it can be shown from \eqref{eq:CFCSI} and \eqref{eq:erg_cap_part_rel_sel1} that the ergodic capacity for full and partial CSI
based relay selection will be
\begin{align}
\label{eq:Cf}
C_F=&e^{1/\bar{\gamma}}\text{log}_2e E_1(1/\bar{\gamma})+\frac{e^{2/\bar{\gamma}}}{2}\text{log}_2e E_1(2/\bar{\gamma})\nonumber\\
&-\frac{2e^{3/\bar{\gamma}}}{3}\text{log}_2e E_1(3/\bar{\gamma})
\end{align}
and
\begin{align}
\label{eq:Cp}
C_P=e^{1/\bar{\gamma}}\text{log}_2e E_1(1/\bar{\gamma})-\frac{e^{3/\bar{\gamma}}}{4}\text{log}_2e E_1(3/\bar{\gamma}),
\end{align}
respectively.
The capacity gain of the full CSI based link selection scheme as compared to the partial CSI based selection scheme can be obtained from \eqref{eq:Cf} and 
\eqref{eq:Cp} is given as
\begin{align}
\label{eq:Capgain}
\Delta C=\frac{e^{2/\bar{\gamma}}}{2}\text{log}_2e E_1(2/\bar{\gamma})-\frac{5e^{3/\bar{\gamma}}}{12}\text{log}_2e E_1(3/\bar{\gamma}).
\end{align}
We have plotted analytical and simulated values of the ergodic capacity for full and partial CSI based selection scheme with a single relay and i.i.d. links in Fig.~\ref{fig:ray2}. It can be seen from Fig.~\ref{fig:ray2} that simulated values closely follow the analytical values of the capacity obtained from \eqref{eq:Cf} and \eqref{eq:Cp}. Further, it can be verified from Fig.~\ref{fig:ray2} that the partial CSI based link selection works poorer than the full CSI based selection scheme in i.i.d. Rayleigh channels specially at high SNRs. Moreover, the partial CSI based selection scheme looses close to 1 bit/sec/Hz as compared to the full CSI based selection scheme at 25~dB SNR as observed after plotting \eqref{eq:Capgain} in Fig.~\ref{fig:ray2}. 

At very high SNR, $2/\bar{\gamma}\approx 3/\bar{\gamma}$, $e^{2/\bar{\gamma}}\approx e^{3/\bar{\gamma}}\approx 1$; therefore, by using the following approximation $E_1(x)\approx e^{-x}\text{ln}\left(1+1/x\right)$~\cite{grand07}, we have
\begin{align}
\label{eq:Capgain1}
\Delta C\approx \frac{\text{log}_2e}{12} \text{ln}\left(1+\bar{\gamma}/2\right).
\end{align}
It can be seen from \eqref{eq:Capgain1} that the capacity gain of the full CSI based scheme monotonically increases with average SNR as compared to the partial CSI based scheme when all links have high SNR. 
\subsection{Comparison of Full and Partial CSI Based Link Selection Schemes in I.N.I.D. Links}
\begin{figure}[t!]\vspace*{-1.0em}
  \begin{center}\hspace*{-1.4em}
    \psfig{figure=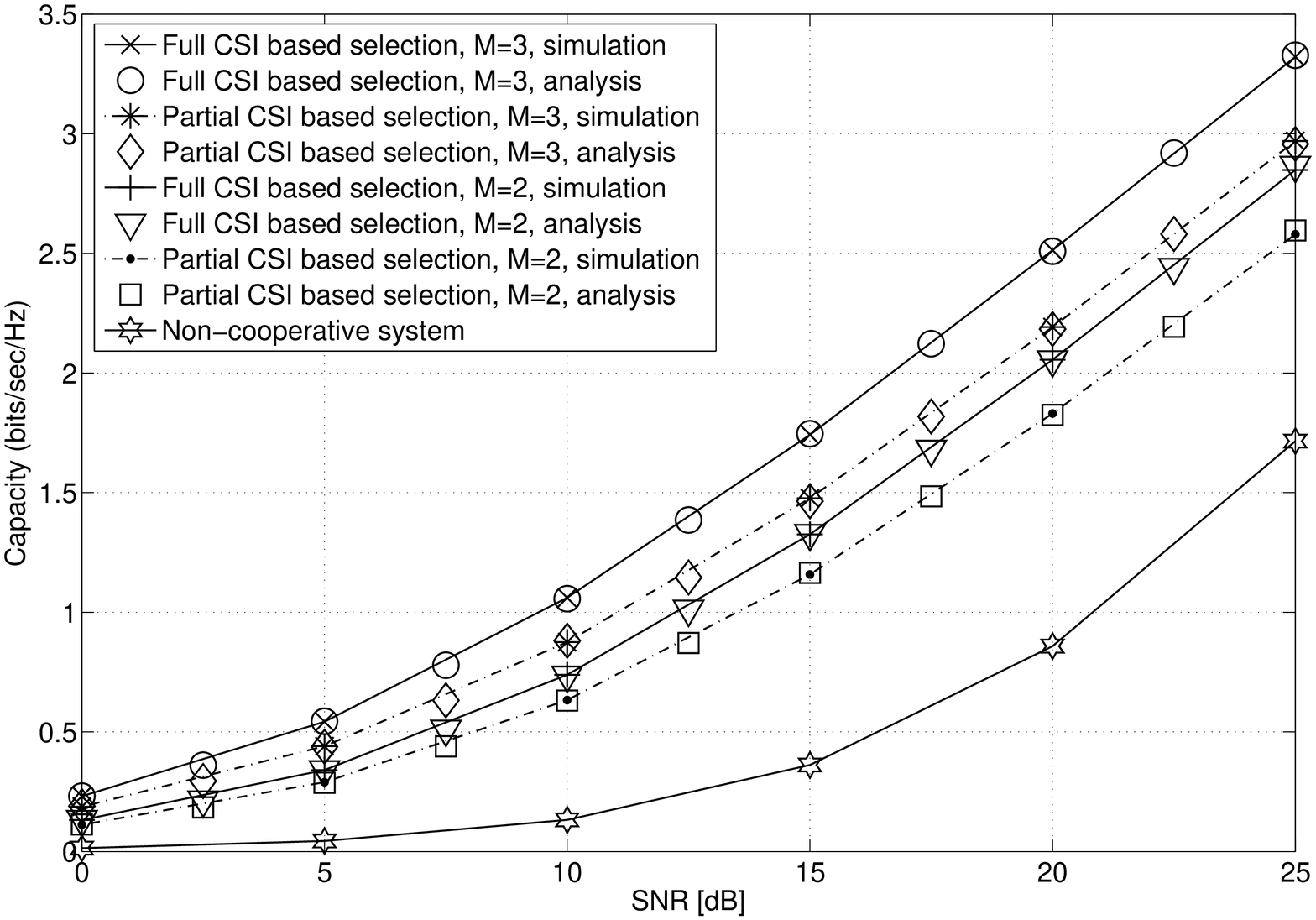,height=3.2in,width=4.0in}
    \vspace*{0em}
    \mycaption{Analytical and simulated ergodic capacity of full and partial CSI based link selection schemes in the two-hop DF system with different number of relays and i.n.i.d. links.}
    \label{fig:rayl}
    \vspace*{-0.0em}
  \end{center}
\end{figure}

In Fig.~\ref{fig:rayl}, we have plotted the simulated and analytical ergodic capacity (bits/sec/Hz) of the full and partial CSI based transmission link selection schemes in a two-hop DF cooperative system with $M=2,3$ relays and direct link between S and D over i.n.i.d Rayleigh fading links. In order to realize the i.n.i.d. channels, it is assumed that $\bar{\gamma}_{ik}=\bar{\gamma}/i$, $\bar{\gamma}_{ki}=\bar{\gamma}/(2i)$, and $\bar{\gamma}_{0}=\bar{\gamma}/100$, where $\bar{\gamma}$ denotes the average SNR and shown on the x-axis of Fig.~\ref{fig:rayl}. The analytical values of the capacity are obtained from \eqref{eq:CFCSI} and \eqref{eq:erg_cap_part_rel_sel1}. It can be seen from Fig.~\ref{fig:rayl} that the simulated capacity results closely match with the analytical values at all SNRs. Further, the full CSI based selection scheme provides better capacity than the partial CSI based link selection scheme at all SNRs considered in the figure. In Fig.~\ref{fig:rayl}, we have also shown the simulated capacity of a non-cooperative system that always uses the direct link for transmission of the source's data to the destination. It can be seen from Fig.~\ref{fig:rayl} that both selection schemes provide significant improvement in the average capacity as compared to the non-cooperative system; capacity is improved with increasing number of relays.  
\section{Conclusions}
We have derived the ergodic capacity of two transmission link selection schemes in a DF cooperative system with multiple relays and a direct link. It has been shown by simulation and analysis that full CSI based selection scheme provides better capacity compared to the partial CSI based selection scheme. Moreover, both selection schemes render better capacity than a non-cooperative system. 
\appendices
\section{Derivation of \eqref{eq:CFCSI}}
\label{app:1}
Let us first consider the first integral term in \eqref{eq:erg_cap} as
\begin{align}
\label{eq:Ai}
A_i&=\int^\infty_{0}\text{log}_2\left(1+x\right)f_{\gamma_i}(x)\int^x_0f_{\gamma^{(i-)}_M}(y)\nonumber\\
&\times \int^x_0f_{\gamma_0}(z)dz dy dx.
\end{align}
We can simplify $A_i$ from \eqref{eq:Ai} as
\begin{align}
\label{eq:Ai0}
A_i&=\int^\infty_{0}\text{log}_2\left(1+x\right)f_{\gamma_i}(x)F_{\gamma^{(i-)}_M}(x)
F_{\gamma_0}(x)dx.
\end{align}
By observing that $F_{\gamma_0}(x)=1-e^{-x/\bar{\gamma}_0}$, using \eqref{eq:CDFmax2} in \eqref{eq:Ai0}, and after some algebra, we get 
\begin{align}
\label{eq:Ai1}
A_i=&\frac{1}{\bar{\gamma}_i}\int^\infty_{0}\text{log}_2\left(1+x\right)e^{-x/\bar{\gamma}_i}\nonumber\\
&\times\Big(1-\sum^{M-1}_{m=0}(-1)^m\mathbb{S}^{(i-)}_{0,m}e^{-x/\bar{\gamma}''_m}\Big)dx. 
\end{align}
Using substitution of variable $t=1+x$ and~\cite[Eq.~(4.331.2)]{grand07} in \eqref{eq:Ai1} results into
\begin{align}
\label{eq:Ai2}
A_i&= e^{1/\bar{\gamma}_i}\text{log}_2e E_1\left(\frac{1}{\bar{\gamma}_i}\right)-\frac{\text{log}_2e}{\bar{\gamma}_i}\sum^{M-1}_{m=0}(-1)^k\mathbb{S}^{(i-)}_{0,m}\nonumber\\
&\times\left(\frac{1}{\bar{\gamma}_i}+\frac{1}{\bar{\gamma}''_m}\right)^{-1} e^{1/\bar{\gamma}_i+1/\bar{\gamma}''_m}E_1\left(\frac{1}{\bar{\gamma}_i}+\frac{1}{\bar{\gamma}''_m}\right). 
\end{align}
The second integral in \eqref{eq:erg_cap} can be written as
\begin{align}
\label{eq:B}
B=\int^\infty_{0}\text{log}_2\left(1+x\right)f_{\gamma_0}(x)F_{\gamma_M}(x)dx.
\end{align}
We can solve the integral of $B$ by using \eqref{eq:CDFmax1}, substitution $t=1+x$, and~\cite[Eq.~(4.331.2)]{grand07}. 
\section{Derivation of \eqref{eq:erg_cap_part_rel_sel1}}
\label{app:2}
The first term of \eqref{eq:erg_cap_part_rel_sel} can be written as
\begin{align}
\label{eq:Ui}
U_i&=\int^\infty_{0}\text{log}_2\left(1+x\right)f_{\gamma_{ik}}(x)\int^\infty_x f_{\Omega_i}(y) F_{\Omega^{(i-)}_M}(y) dy dx.
\end{align}
From \eqref{eq:cdf_omegaM} and \eqref{eq:Ui}, and after some algebra, we get
\begin{align}
\label{eq:Ui0}
&U_i=\frac{\text{log}_2e}{\bar{\gamma}_{ik}}\int^\infty_{0}\text{log}_2\left(1+x\right)e^{-x/\bar{\gamma}_i}dx-\frac{\text{log}_2e}{\bar{\gamma}_{ik}\bar{\gamma}_{ki}}\nonumber\\
&\times\sum^{M-1}_{m=0}(-1)^m \mathbb{S}^{(i-)}_{0,m}\left(\frac{1}{\bar{\gamma}_{ki}}+\frac{1}{\bar{\Omega}'_m}\right)^{-1}\int^\infty_{0}\text{log}_2\left(1+x\right)\nonumber\\
&\times e^{-\left(\frac{1}{\bar{\gamma}_i}+\frac{1}{\bar{\Omega}'_m}\right)x}dx.
\end{align}
By substitution of variable $t=1+x$ and using~\cite[Eq.~(4.331.2)]{grand07} in \eqref{eq:Ui0}, results into 
\begin{align}
\label{eq:Ui1}
U_i&=\frac{\bar{\gamma}_i}{\bar{\gamma}_{ik}}e^{1/\bar{\gamma}_i}\text{log}_2e E_1\left(\frac{1}{\bar{\gamma}_i}\right)-\frac{\text{log}_2e}{\bar{\gamma}_{ik}\bar{\gamma}_{ki}}\sum^{M-1}_{m=0}(-1)^m\mathbb{S}^{(i-)}_{0,m}\nonumber\\
&\times\left(\frac{1}{\bar{\gamma}_i}+\frac{1}{\bar{\Omega}'_m}\right)^{-1}\left(\frac{1}{\bar{\gamma}_{ki}}+\frac{1}{\bar{\Omega}'_m}\right)^{-1} e^{1/\bar{\gamma}_i+1/\bar{\Omega}'_m}\nonumber\\
&\times E_1\left(\frac{1}{\bar{\gamma}_i}+\frac{1}{\bar{\Omega}'_m}\right).
\end{align}
The remaining integrals in \eqref{eq:erg_cap_part_rel_sel} can be solved by using \eqref{eq:cdf_omegaM}, substitution of variable method, and~\cite[Eq.~(4.331.2)]{grand07}.   
\bibliography{IEEEabrv,biblitt}
\bibliographystyle{IEEEtran}
\end{document}